*Article*

# A Congestion Control System Based on VANET for Small Length Roads

**Ruchin Jain**[1, *]

[1]School of Computer Studies, AMA International University BAHRAIN (AMAIUB), Bahrain
r.jain@amaiu.edu.bh
*Correspondence: r.jain@amaiu.edu.bh

**Abstract:** As vehicle population has been increasing on a daily basis, this leads towards increased number of accidents. To overcome this issue, Vehicular Ad Hoc Network (VANET) has come up with lot of novel ideas such as vehicular communication, navigation and traffic controlling. In this study, the main focus is on congestion control at the intersections which result from unclear ahead. For this purpose, a city lane and intersection model has been proposed to manage vehicle mobility. It shows the actual vehicle to vehicle and vehicle to traffic infrastructure communication. The experiment was conducted using Network Simulator 2 (NS 2). The implementation required modelling the road side unit, traffic control unit, and on-board unit along the roadside. In the simulation, including traffic volume, the distance between two signals, end-to-end delay, packet delivery ratio, throughput and packet lost were taken into consideration. These parameters ensure efficient communication between the traffic signals. This results in improved congestion control and road safety, since the vehicles will be signalled not to enter the junction box and information about other vehicles.

## 1. Introduction

Vehicular Ad Hoc Network (VANET) is a potential area in road safety, traffic management, and information passing for the drivers and commuters. VANETs are getting a major amount of focus as it can provide a large number of services. Many VANETs have been proposed by other researchers. However, this study suggests a novel solution to achieve congestion free traffic at the intersection to avoid the vehicles to enter the yellow box area at the intersection. The objective in the design of such networking is to achieve a congestion free traffic control to avoid the vehicles by using vehicle-to vehicle, vehicle to infrastructure and vehicle-to-roadside unit wireless communication. The On Board Units (OBU) vehicles were used to communicate with each other as well as with Road Side Units (RSUs) located at main points on the road at a calculated interval. A network is created by the connection of the vehicles and RSUs, called a vehicular ad hoc network (VANET), further, the RSUs connected to internet. In this study a VANET based vehicle to vehicle and vehicle to signal control that is, the infrastructure based road traffic control system which can collect traffic information from individual cars and disseminate information of road traffic over a network to control the traffic signalling is presented so as to avoid the vehicle from the entering the yellow box on the intersection.

In the current road traffic signalling systems road sensors are used in some busy streets of cities to estimate traffic arrivals which provide a very limited information about the traffic status [2].

## 2. Literature Review

Many proven ways and means were suggested by different researchers to trounce the issues in managing the traffic. Researcher further carried on the studies on collision warning algorithms for individual vehicle drivers for the purpose safety and designing the system. Furthermore, it deals with the improvement of safety systems in the area of Vehicular Ad Hoc Networks. In the past few years, collision warning system have been developed to help reduce the rear-end collision. However, these types of systems dependent on the vehicles drivers, thereby the distinct characteristics of individual drivers are hereby ignored [3].

The impact of 802.11p channel hopping on VANET communication protocols was narrated by E. A. Donato and G. Maia. VANETs are a particular type of a moving networks of vehicles as the nodes consisting of storage, processing and wireless communication capabilities. The wireless access in vehicular environments presents architecture, based on a division into multiple channels. Each channel uses a switching mechanism for the selection of channels, since only one channel is active at a given time. However, in some scenarios, collisions can be a result of this channel switching mechanism approach used in wave introduces an undesirable effect that allows different vehicles to transmit simultaneously [4].

## 3. VNET Traffic Control System

The VANET categorized particularly for the short range communication among the mobile host vehicles as well as between the vehicles and the road side information infrastructure. Generally, the moving vehicles are equipped with On Board Unit (OBU) and the road side communication infrastructures are referred to Road Side Units (RSUs) which is further used to control the traffic signal at the intersection. The new WLAN standard IEEE 802.11 referred to as Dedicated Short Range Communication (DSRC) which forms the bases for the V2V and the V2I communication as well is used to access the wireless in the vehicular environment. The licensed spectrum of 75MHz has been allocated at 5.9GHz for the DSRC. The IEEE 802.11a standard is similar to the physical layer of the IEEE 802.11p. Following figure shows the overview of Vehicular Ad Hoc Networks (VANET).

Also, TCC supports to identify the location of the destination vehicle in a quick manner. Comparing to traditional VANET, better performance of the BUS-VANET can be obtained by higher delivery rate and less delivery delays [5].

U. Kumaran and Dr. R. S. Shaji researched that on vertical handover in vehicular ad hoc network using multiple parameters.  Mainly because of resource management, Number of handover is the main parameter considered. Unnecessary handover reduces throughput and network occupancies performance.  However, in the vertical handover decision algorithm implementation of multi-criteria decision making of a vehicular ad hoc network increases the network performance in view of load balance index and the number of hand off [6]. The proposed system is shown in Figure 1.

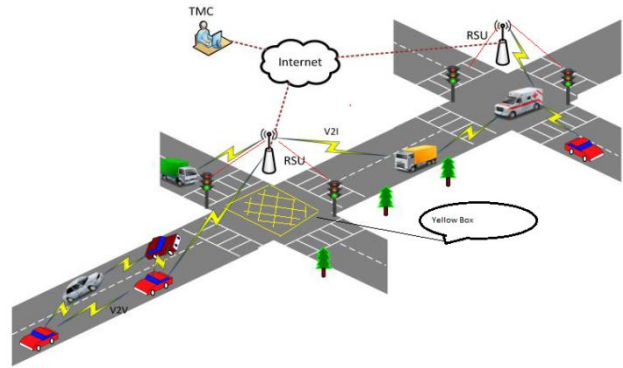

**Figure 1**. VANET in action for Congestion Control at Intersection.

## 4. Methodology

In this research, a basic traffic control system for traffic light based on the use of 802.11p standard is proposed. The target is to send a message to a vehicle about not entering the yellow box at the intersection with high reliability and low delay and to control the traffic signalling. The performance of this proposed system was also evaluated. Issues related to traffic safety such as the propagation delay of warning messages in the vicinity, the number of blind vehicles, and the total number of packets received by each vehicle is addressed.

## 5. Simulation

For this study, the performance simulation parameters that are taken into account are shown in Table 1.

**Table 1.** Parameters Used for Performance Evaluation.

| Simulation Parameter | Value |
|---|---|
| **Number of vehicle (nodes)** | 200 |
| **Maximum speed of vehicle** | 5 – 20 m/sec |
| **Scenario size** | 800 x 800 m |
| **Length of the vehicle** | 4.5 m |
| **Number of vehicle crossed on green signal** | 175 |
| **Distance between two traffic signal** | 50m |
| **Normal Packet size** | 512B |
| **Warning Packet size** | 256B |
| **Simulation time** | 5 minutes |
| **Sensor transmission range** | 75 m |
| **Vehicle transmission range** | 200 m |

**Table 2.** Varying Parameters.

| Parameters | Data Rate | Packet Size | No. of Vehicles |
|---|---|---|---|
| **Packet Delivery ratio (%)** | 97% - 98% | 99% - 99.5% | 98% - 99% |
| **Packet Loss (Bytes)** | < 4 | < 5 | < 2 |
| **End-to-End Delay (ms)** | < 25 ms | < 25 ms | < 6.2 ms |
| **Throughput (Kbps or Mbps)** | 1.5 Mb/s | 150 Kb/s | 15-100 Kb/s |

Table 2 refers the Quality of Service (QoS) [7-9] parameters of the network considered for our evaluation. The Packet Delivery Ratio (PDR) is the ratio of the number of delivered data packet to destination. Higher the value of PDR the better performance of the protocol. End-to-End Delay is the average taken by data packets to arrive in the destination which is lower. Packet loss is total number of packets dropped during the simulation and the amount of data received at destination is given by throughput values.

**6. Concluding Discussions**

The proposed system promises improved performance measures by attaining reduced delay and higher Packet Delivery Ratio which are vital parameters considered in the study. The analysis of the simulation results reveals that there are lesser congestions at the yellow box to minimize the traffic jams at the intersection junctions. Thus the results evidence that the VANET improved traffic efficiency, resulting in decreased travel times and safety concerns. Since positive results were achieved by simulation of this VANET based scenario, the future direction of the research is to prototype a Cloud Computing [10-11] based IoT [12-13] system to experiment this proposed system in real traffic.

**References**

[1] Kumar, A., Sinha, M., "Overview on Vehicular ad hoc network & its security issues", computing for sustainable Global Development (INDIACom) 2014 International congerence on, vol., no., pp. 792, 797, 5-7 march 2014

[2] B. Zhou, J. Cao, H. Wu;, "Adaptive traffic light control of multiple intersections in WSN based ITS", 978-1-4244-8331, IEEE, 2011.

[3] Rakhshan, A., Pishor-Nik, H., Fisher, D., Nekoui, M., "Tuning collision warning algorithms to individual drivers for design of active safety systems," Globalcom Workshops (GC workshops), 2013 IEEE, vol., no., pp. 1333, 1337, 9-13 Dec 2013

[4] Donato, E.A.; Maia Menezes, J.G.; Madeira, E.R.M.; Villas, L.A., "Impact of 802.11p channel Hopping on VANET communication protocols," Latin America Transactions, IEEE (Revista IEEE America Latina), vol. 13, no.1, pp. 315, 320, Jan. 2015 doi: 10.1109/TLA.2015.7040664

[5] Xiaoxiao Jiang; David H.C. Du, "Bus-VANET: A bus Vehicular network Integrated with Traffic Infrastructure", IEEE intelligent transportation system magazine, 1939-1390/15, 2015 IEEE

[6] U. Kumaran; Dr. R. S. Shaji., "Vertical handover in Vehicular Ad-hoc Network using Multiple Parameters," 2014 International Conference on Control,

[7] Mahdi H. Miraz, Muzafar A. Ganie, Maaruf Ali, Suhail A. Molvi, and AbdelRahman H. Hussein, "Performance Evaluation of VoIP QoS parameters using WiFi-UMTS networks" in the "Transactions on Engineering Technologies", Chapter 38, DOI: 10.1007/978-94-017-9804-4_38, Print ISBN: 978-94-017-9803-7, Online ISBN: 978-94-017-9804-4, pp 547-561, 2015, Springer-Verlag, Available: https://link.springer.com/chapter/10.1007/978-94-017-9804-4_38.

[8] Mahdi H. Miraz, Suhail A. Molvi, Maaruf Ali, Muzafar A. Ganie, and AbdelRahman H. Hussein, "Analysis of QoS of VoIP Traffic through WiFi-UMTS Networks" in the proceedings of the World Congress on Engineering (WCE 2014), held at Imperial College, London, UK, 2-4 July 2014, ISBN-13: 978-988-19252-7-5, Print ISSN: 2078-0958, Online ISSN: 2078-0966, Vol. 1, pp. 684-689, Available: http://www.iaeng.org/publication/WCE2014/WCE2014_pp684-689.pdf.

[9] Mahdi H. Miraz, Muzafar A. Ganie, Suhail A. Molvi, Maaruf Ali, and AbdelRahman H. Hussein, "Simulation and Analysis of Quality of Service (QoS) Parameters of Voice over IP (VoIP) Traffic through Heterogeneous Networks" in the International Journal of Advanced Computer Science and Applications (IJACSA), Online ISSN: 2156-5570, Print ISSN: 2158-107X, Volume 8, No 7, July 2017, pp. 242-248, published by Science and Information (SAI)

Organization, DOI: 10.14569/IJACSA.2017.080732, Available: http://thesai.org/Publications/ViewPaper?Volume=8&Issue=7&Code=ijacsa&SerialNo=32 .

[10] Maaruf Ali and Mahdi H. Miraz, "Recent Advances in Cloud Computing Applications and Services", International Journal on Cloud Computing (IJCC), ISSN 1982-9445, vol. 1, no. 1, pp. 1-12, February 2014, DOI: 10.1007/CC8554885.14.01.

[11] Maaruf Ali and Mahdi H. Miraz, "Cloud Computing Applications", in the Proceedings of the International Conference on Cloud Computing and eGovernance - ICCCEG 2013 held at Internet City, Dubai, United Arab Emirates, 19-21 June 2013, pp. 1-8, Ed. Manikandan Ayappan, published by Assoc. of Scientists, Developers and Faculties, ISBN: 978-81-925233-2-3, DOI: 10.ASDFOI/925233.001.

[12] Mahdi H. Miraz, Maaruf Ali, Peter Excell and Rich Picking, "A Review on Internet of Things (IoT), Internet of Everything (IoE) and Internet of Nano Things (IoNT)" in the proceedings of the fifth international IEEE conference on Internet Technologies and Applications (ITA 15), held at Glyndwr University, Wrexham, North East Wales, UK, DOI: 10.1109/ITechA.2015.7317398, Print ISBN: 978-1-4799-8036-9, pp 219-224, 8-11 September 2015, Published by IEEE, Available: http://ieeexplore.ieee.org/document/7317398/?arnumber=7317398.

[13] Zainab Alansari, Safeeullah Soomro, Mohammad Riyaz Belgaum, Shahaboddin Shamshirband. (2016). "The Rise of Internet of Things (IoT) in Big Healthcare Data: Review and Open Research Issues" International Conference on Advanced Computing and Intelligent Engineering (ICACIE2016), India. Advances in Intelligent Systems and Computing (AISC) series of Springer, DOI: 10.1007/978-981-10-6875-1_66, Available: https://link.springer.com/chapter/10.1007/978-981-10-6875-1_66.